\documentclass[12pt]{JHEP3}
\usepackage{epsfig}

\newcommand{\im}{{\rm Im}}
\newcommand{\re}{{\rm Re}}
\newcommand{\be}{\begin{equation}}
\newcommand{\ee}{\end{equation}}
\newcommand{\bea}{\begin{eqnarray}}
\newcommand{\eea}{\end{eqnarray}}
\newcommand{\beq}{\begin{equation}}
\newcommand{\eeq}{\end{equation}}
\newcommand{\ba}{\begin{array}}
\newcommand{\ea}{\end{array}}
\newcommand{\beqa}{\begin{eqnarray}}
\newcommand{\eeqa}{\end{eqnarray}}

\newcommand{\cL}{{\cal L}}
\newcommand{\cH}{{\cal H}}
\newcommand{\cA}{{\cal A}}

\newcommand{\cO}{{\cal O}}
\newcommand{\BR}{{\cal B}}
\newcommand{\da}{^\dagger}
\newcommand{\no}{\nonumber}
\newcommand{\nnu}{\nonumber}
\newcommand{\lsim}{\stackrel{<}{_\sim}}
\newcommand{\gsim}{\stackrel{>}{_\sim}}

\newcommand{\dbar}{{\bar d}}

\newcommand{\lbar}{{\bar \ell}}
\newcommand{\tu}{{\tilde u}}
\newcommand{\ttop}{{\tilde t}}
\newcommand{\tq}{{\tilde q}}

\def \vckm{V}
\def \text{\mathrm}

\def\arnps#1#2#3{  {\it Annu. Rev. Nucl. Part. Sci. }{\bf #1} (#2) #3}
\def\npb#1#2#3{    {\it Nucl. Phys. }{\bf B #1} (#2) #3}
\def\plb#1#2#3{    {\it Phys. Lett. }{\bf B #1} (#2) #3}

\def\prd#1#2#3{    {\it Phys. Rev. }{\bf D #1} (#2) #3}

\def\prl#1#2#3{    {\it Phys. Rev. Lett. }{\bf #1} (#2) #3}

\def\rmp#1#2#3{    {\it Rev. Mod. Phys. }{\bf #1} (#2) #3}

\def\ibid#1#2#3{   {\it }{\bf #1} (#2) #3}
\def\jhep#1#2#3{   {\it JHEP  }{\bf #1} (#2) #3}

\title{$B_{s,d} \to \ell^+\ell^-$ and $K_L \to \ell^+\ell^-$ in SUSY models
           with non-minimal sources of flavour mixing } 
\author{Gino Isidori \\ Theory Division, CERN, CH-1211 Geneva 23, Switzerland \\
 INFN, Laboratori Nazionali di Frascati, Via E. Fermi 40, I-00044 Frascati, Italy \\
 E-mail: \email{Gino.Isidori@cern.ch} }
 \author{Alessandra Retico \\ Theory Division, CERN, CH-1211 Geneva 23, Switzerland \\
 INFN, Sezione di Roma and Dipartimento di Fisica, \\
           Universit\`a di Roma ``La Sapienza'', P.le A. Moro 5, I-00185 Rome, Italy \\
 E-mail: \email{Alessandra.Retico@cern.ch} }

\abstract{
We present a general analysis of $B_{s,d} \to \ell^+\ell^-$ and  
$K_L \to \ell^+\ell^-$ decays 
in supersymmetric models with non-minimal sources of flavour mixing. 
In spite of the existing constraints on off-diagonal 
squark mass terms, these modes could still receive sizeable corrections, 
mainly because of Higgs-mediated FCNCs arising at large $\tan\beta$. 
The severe limits on scenarios with large $\tan\beta$ 
and non-negligible $\tilde d^i_{R(L)}$--$\tilde d^j_{R(L)}$ mixing
imposed by the present experimental bounds on these modes 
and $\Delta B=2$ observables are discussed in detail. 
In particular, we show that scalar-current contributions to 
$K_L \to \ell^+\ell^-$ and $B$--$\bar B$ mixing 
set non-trivial constraints on the possibility 
that $B_s \to \ell^+\ell^-$ and $B_d \to \ell^+\ell^-$ receive 
large corrections.}

\preprint{CERN--TH/2002--192}


\begin{document}

\section{Introduction}
The rare dilepton decays 
$B_{s,d} \to \ell^+\ell^-$ and $K_L \to \ell^+\ell^-$ 
are ideal candidates to deeply investigate  
the dynamics of quark-flavour mixing. 
On the one side, they are highly sensitive 
to physics beyond the Standard Model (SM),
like all processes mediated by flavour-changing 
neutral-current (FCNC) amplitudes. 
On the other side, the helicity suppression of vector 
contributions gives to these modes a unique sensitivity
to scalar currents, which are typically 
negligible in other rare decays. 

In this paper we analyse the rare dilepton decays 
in supersymmetric (SUSY) models with flavour 
non-universal soft-breaking terms. As is well known, 
this class of models is strongly constrained by 
currently available data on $\Delta F=2$ and 
$\Delta F=1$ amplitudes (see e.g. Ref.~\cite{GGMS}).
However, a large fraction of the parameter space 
still remains to be explored. The only 
conclusion that can be drawn at present is 
that the possible new sources of flavour mixing, 
if any, must have a rather non-trivial structure. 
For instance, as emphasized recently in Ref.~\cite{CMM},
the possibility of a large  $\tilde b_{R}$--$\tilde s_{R}$
mixing, motivated by SO(10) unification
and neutrino data, is still open.

Supersymmetric contributions to rare FCNC decays have 
been widely discussed in the literature. However, 
most of the analyses are focused on scenarios 
with $\tan\beta = O(1)$. As we shall show, 
if $\tan\beta = O(1)$ there is little to learn 
from the dilepton modes, which cannot be extracted 
elsewhere. On the other hand, 
things change substantially at large $\tan\beta$,
because of the scalar FCNCs induced by 
the effective coupling of $H^*_U$ to 
down-type quarks (the so-called 
non-holomorphic terms) \cite{Babu,IR}. 

As discussed in Refs.~\cite{Babu}--\cite{Nierste}, at large $\tan\beta$
the rates of $B_{s,d} \to \ell^+\ell^-$ decays could be substantially 
enhanced with respect to the SM even without 
new sources of flavour mixing. As pointed out in \cite{IR},
the same conclusion does not hold for $K_L \to \ell^+\ell^-$.
In this paper we analyse how the situation 
is modified by non-minimal flavour structures. 
Interestingly enough, in this case also  
$K_L \to \ell^+\ell^-$ play an important role,
setting stringent bounds on off-diagonal squark 
mass terms. 

The issue of SUSY contributions to $B_{s,d} \to \ell^+\ell^-$
decays in the presence of non-universal squark mass terms has
already been addressed in Refs.~\cite{Chankowski,Bobeth2}.
Here we extend these works by discussing all possible 
$\tan\beta$-enhanced terms, both those arising at higher orders
and those not related to scalar FCNCs. We also clarify the issue 
of the diagonalization of the quark mass matrix 
in the presence of non-universal non-holomorphic terms.
Last but not least, in our analysis we  
take into account the constraints from scalar-current 
contributions to $K_L \to \ell^+\ell^-$ and $B$--$\bar B$ mixing,
which have not been discussed before. 

\section{Generalities}
The part of the $\Delta F=1$ effective Hamiltonian
relevant to  $P[\dbar^i d^j]\to l^+l^-$ decays is
\bea
\!\!\!\! \cH^{\rm eff} 
%=-\frac{G_F}{\sqrt{2}}\frac{\alpha}{\pi\sin^2\theta_W}
=- \frac{G_F^2 M_W^2}{\pi^2}\,
%V_{tb}^{}V_{tq}^{*} 
[C_A \cO_A + C_A' \cO_A' +
C_S \cO_S  + C_P \cO_P + C_S' \cO_S' + C_P' \cO_P']{\rm~+\ h.c.},
\label{eq:Heff1}
\eea
where 
\beqa
&& \cO_A  = \dbar_L^i \gamma^\mu   d^j_L \lbar \gamma_\mu\gamma_5
\ell~, \qquad
\cO_A'  = \dbar_R^i\gamma^\mu d^j_R \lbar \gamma_\mu\gamma_5
\ell~,\no \\ 
&& \cO_S   = \dbar_R^i d^j_L \lbar \ell~, \qquad\qquad\quad
\cO_S'  = \dbar_L^i d^j_R \lbar \ell~,  \no\\
&& \cO_P   =  \dbar_R^i d^j_L \lbar \gamma_5 \ell~, \qquad\qquad 
\cO_P'  =  \dbar_L^i d^j_R \lbar \gamma_5 \ell~.
\label{eq:sc_ops}
\eeqa
According to this normalization of the effective operators, 
writing hadronic matrix elements of axial and pseudoscalar currents as 
\beq
\langle 0 |\bar q^i \gamma_\mu\gamma_5  q^j | P(p)\rangle = i p_\mu f_{P}~, 
\qquad \langle 0 |\bar q^i \gamma_5  q^j | P(p)\rangle =  -i f_{P} 
\frac{M^2_{P } }{ (m_i +m_j )}~, 
\label{eq:fbdef}
\eeq
the most general expression of the short-distance (SD) contribution 
to  $P[\dbar^i d^j]\to l^+l^-$ branching ratios reads
\bea
&& \BR_{\rm SD} (P \to \ell^+ \ell^-) 
   =\frac{G_F^4 M_W^4}{8\pi^5} 
    f^2_{P } M_{P } \tau_{P } m^2_\ell 
    \left(1-\frac{4 m^2_\ell}{M^2_{P }}\right)^{1/2} \times \nnu\\
&& \qquad \left[  \left(1-\frac{4 m^2_\ell}{M^2_{P }}\right) 
       \left| \frac{M^2_{P } ( C_S-C_S')  }{2 m_\ell (m_i+m_j ) } \right|^2 \right.
    \left.  +\left| \frac{M^2_{P}(C_P-C_P')}{2m_\ell (m_i+m_j) }
       + C_{A}-C_{A}' \right|^2\right]~,
\label{eq:Br}
\eea
where the Wilson coefficients are understood 
to be evaluated at a scale of $O(M_P^2)$.
In $B_{s,d}$ modes this is the only relevant 
contribution to the branching ratios, 
whereas in $K_L$ decays also long-distance 
effects have to be taken into account \cite{KL}.
 
Within the SM only $\cO_A$ has a non-vanishing Wilson coefficient, 
\beq
C^{\rm SM}_{A} (M_W^2) =  - V_{3 i}^{*} V_{3 j}^{}~Y_0\left( \frac{m^2_t}{M^2_W} \right) 
 -  V_{2 i}^{*} V_{2 j}^{}~Y_0\left( \frac{m^2_c}{M^2_W} \right)~, 
\eeq
where the charm contribution, which receives sizeable 
corrections in the running down to the meson scale, 
plays a non-negligible role only in the $K_L$ case \cite{BBL}. 
In the SUSY scenarios we are considering in this paper, 
all the $C_i$ could have, in principle, a relevant impact 
on the decay amplitudes. 

We stress that, 
since the ratios $(C_{S,P}-C'_{S,P})/m_\ell$ are independent of $m_\ell$,
to a first approximation the relative weight of the various contributions is 
independent of $m_\ell$, whereas the overall branching ratios scale 
like $m_\ell^2$. This implies that i) $e^+e^-$ modes are never interesting;
ii) the difficult detection of $\BR(B_{s,d} \to \tau^+\tau^-)$ with respect to 
$\BR(B_{s,d} \to \mu^+\mu^-)$ is partially compensated 
by an enhancement factor $\sim (m_\tau/m_\mu)^2$.

\section{Axial operators}
\label{sec:axial}
The contributions to the axial operators $\cO_A$ and 
$\cO_A'$ can be divided into two big categories: $Z$ penguins
and box diagrams. Those in the former class receive contributions 
from charginos,
neutralinos and gluinos, while gluino exchange does not contribute 
to the latter class. 

As discussed by several authors \cite{Kpnn1}--\cite{Lunghi}, 
box diagrams always play a subdominant role: in the case of 
$\tilde q^{i}_L$--$\tilde q^j_R$ mixing the $Z$ penguins
are by far dominant, in the case of $SU(2)_L$-conserving 
mixing ($\tilde q^{i}_L$--$\tilde q^j_L$ or $\tilde q^{i}_R$--$\tilde q^j_R$)
the constraints from $\Delta F=2$ processes 
forbid sizeable effects.

$Z$-penguin diagrams correspond to dimension-six effective 
operators of the type $\bar q^i_{L(R)} \gamma^\mu q_{L(R)}^j 
H^\dagger D_\mu H$, 
where $H$ denotes the Higgs field. For this reason,
to be non-vanishing, these diagrams require a 
double chirality flip, which can arise either by 
the $\tilde q^{i}_L$--$\tilde q^j_R$ mixing or by 
the gaugino--Higgsino mixing (for charginos and neutralinos only).
Also in this case, $\Delta F=2$ constraints  
forbid sizeable effects in the case of $SU(2)_L$-conserving 
squark mixing. As a consequence, possible large effects 
(with respect to the SM) are controlled only by the   
$\tilde q^{i}_L$--$\tilde q^j_R$ mixing terms, 
both in the up- and in the down-type sector.

\subsection{Charginos}
\label{sect:charg_1}
Involving up-type squarks, chargino contributions to $Z$ penguins 
are mildly sensitive to the value of $\tan\beta$. These contributions,
which we report here for completeness,  
have been widely discussed in the literature in the context of 
$B (K) \to X_s (X_d) \ell^+ \ell^-$ processes 
\cite{Kpnn1}--\cite{Lunghi}. 
Before any mass expansion, we can write \cite{Colangelo}
\beq 
[C_A]^{Z^0}_{\chi} = {1\over 8 g^2} L^j_{nl}{\bar L}^i_{mk} F_{nmlk}~ 
\label{eq:Wchi}
\eeq
where
\beqa
L^j_{nl} &=& - g {\hat \Gamma}^U_{l j_L}{\hat V}\da_{1n} 
            + y^{u}_q \vckm_{qj}{\hat \Gamma}^U_{l q_R}{\hat V}\da_{2n}~, \no \\
{\bar L}^i_{mk} &=& - g ({\hat \Gamma}^U)\da_{i_L k}{\hat V}_{m1} 
            + y^{u}_q \vckm^*_{qi}({\hat \Gamma}^U)\da_{q_R k}{\hat V}_{m2}~, \no \\
F_{nmlk} &=& {\hat V}_{n1}{\hat V}_{1m}\da~ \delta_{lk}~ k(x_{mk},x_{nk}) 
             - 2 {\hat U}_{m1} {\hat U}\da_{1n}~
\delta_{lk}~ \sqrt{x_{mk}x_{nk}} f(x_{mk},x_{nk}) \nonumber \\
         &&  -\delta_{mn}~ {\hat \Gamma}^U_{k q_L} ({\hat \Gamma}^U)\da_{q_L l}~ k(x_{mk},
x_{lk})~;
\eeqa
$g$ is the $SU(2)_L$ gauge coupling, 
$V$ denotes the CKM matrix,  $y^u_q =\sqrt{2} m_{u_q} /(v \sin\beta)
= g m_{u_q}/(\sqrt{2} M_W \sin\beta)$ and  
the loop functions $k(x,y)$ and $f(x,y)$ can be found in the 
appendix.\footnote{~As usual $x_{\alpha \beta}=M^2_\alpha/M^2_\beta$, with the indices ($n$, $m$) 
and  ($l$, $k$) denoting chargino and squark-mass eigenstates, respectively.}
Here ${\hat V}$ and ${\hat U}$
are the unitary matrices which diagonalize the chargino mass matrix
[${\hat U}^* M_\chi {\hat V}\da = \mbox{diag}(M_{\chi_1},M_{\chi_2})$]
and ${\hat \Gamma}^U$ is the one which diagonalizes the up-squark mass matrix,
written in the basis where the $d_L^i-\tu_L^j-\chi_n$ 
coupling is family diagonal and the $d_L^i-\tu_R^j-\chi_n$ 
one is ruled by the CKM matrix. 

The contribution of chargino $Z$-penguin diagrams  
to $C_{A}'$ necessarily involves two down-type Yukawa couplings 
($y^d_{i} y^d_{j}$), which cannot be both of the third family. 
As a result, even at large $\tan\beta$, this contribution 
is negligible with respect to the one in Eq.~(\ref{eq:Wchi}).

A detailed discussion on the relative weight of the various terms 
appearing in Eq.~(\ref{eq:Wchi}), up to the second order in 
the mass expansion, can be found in \cite{Colangelo,Lunghi}. 
Due to the off-diagonal CKM structure, the zeroth 
order term is not vanishing. However, this is not 
very interesting since it reaches at most $\cO(10\%)$ 
of the SM contribution: the potentially largest effects 
are induced by terms of first or second order in the expansion of 
the squark mass matrix around the diagonal \cite{Colangelo,Lunghi}. 
A substantial simplification arises if, in addition to the squark mass matrix, 
also the chargino mass matrix is perturbatively expanded around the diagonal.
In this case we can write 
\beq
\frac{ [C_A]^{Z^0}_{\chi} }{ C_A^{\rm SM} } \approx
\frac{1}{8} \left[
 y_t   \frac{ (\delta^U_{RL})_{3j}\delta^{^\chi}  }{ V_{3 j} }
 f_1[x_{\alpha \beta}] + 
 y_t \frac{ (\delta^U_{LR})_{i3} \delta^{^\chi} }{ V_{3 i}^* } f_1[x_{\alpha \beta}] 
 +  \frac{ (\delta^U_{LR})_{i3}(\delta^U_{RL})_{3j} }{ V_{3 i}^* V_{3 j} }  f_2[x_{\alpha \beta}]
 \right]+\ldots, 
\label{eq:deltachi}
\eeq
where 
\beq
 (\delta^U_{RL})_{ij} = 
 \frac{(M^2_{\tilde U})_{i_R j_L}}{ \langle M^2_\tq \rangle  }~, 
 \qquad \qquad ~
 \delta^{^\chi} = \frac{ M_W }{ \langle M_\chi \rangle }~,
 \label{eq:deltas}
\eeq
and the dots denote terms not enhanced by 
inverse CKM matrix elements, which can be safely neglected.
The explicit expressions of the 
adimensional functions $f_{1,2}$, which depend on the 
various sparticle mass ratios (and mildly on $\tan\beta$), 
can be extracted from the full result in Eq.~(\ref{eq:Wchi}): 
for $M_{\ttop_R}/M_{\tu_L}\geq 1/2$ and $M_{\chi_n}/M_{\tu_L}\geq 1/3$,
one finds $|f_{1}|\lsim 0.1$ and  $|f_{2}|\lsim 0.4$ (the upper figures 
are obtained for the minimal value of $M_{\ttop_R}$). 

As anticipated, large 
effects are in principle possible in the presence 
of large $\tilde u^{i}_L$--$\tilde u^j_R$ mixing. However, 
both in the kaon and in the $B$ system this possibility has recently 
been severely restricted by new experimental data on the 
non-helicity-suppressed modes. In the kaon case a clean and stringent  
bound is set by $\BR(K^+ \to \pi^+\nu\bar\nu)$ \cite{E787}. Similarly, 
in the $B$ system severe bounds are obtained by the recent data on 
both exclusive and inclusive $B_d \to X_s \ell^+ \ell^-$ transitions 
\cite{Bsll_exp}. As a consequence of these new data, we are led to
the model-independent conclusion that $|C_A/C_A^{\rm SM}| < 2$
in the $s \to d$ channel \cite{Kpnn_DI} and is even closer to unity in the $b\to s$ 
case \cite{Ali}.  
The only channel without strong experimental bounds is the 
$b\to d$ one. However, once the vacuum-stability 
bounds of Ref.~\cite{CD} are imposed on $(\delta^U_{RL})_{31}$,
also in this case one can at most obtain $|C_A/C_A^{\rm SM}| = O(1)$.

\subsection{Gluinos}
Contributions  to $Z$ penguins which involve down-type squarks, 
with gluinos and neutralinos, are very  
sensitive to the value of $\tan\beta$. In particular, at 
small $\tan\beta$ their effect is completely negligible 
because of the smallness of left--right down-type mass 
insertions. This conclusion is not necessarily true at
large $\tan\beta$, where, in principle,  
left--right down-type mass insertions can
become as large as up-type ones. 

The largest effects are expected from gluino exchange, 
whose contribution to $C_A$, before any mass expansion, 
can be written as
\beq 
[C_A]^{Z^0}_{g} = \frac{1}{8 g^2} A^j_{bl}{\bar A}^i_{ak} 
I_{balk}\, , 
\eeq
where
\bea
A^j_{bl}&=& - \sqrt{2} g_s T^b {\hat \Gamma}^D_{l j_L}~, \no \\
{\bar A}^i_{ak}&=& - \sqrt{2} g_s T^a ({\hat \Gamma}^D)\da_{i_L k}~, \no \\
I_{balk} &=& \delta_{ab}~ {\hat \Gamma}^D_{k q_R} 
({\hat \Gamma}^D)\da_{q_R l}~ k(x_{gl}, x_{kl})~, 
\eea
$g_s$ and $T^a$ denote $SU(3)_c$ coupling and generators, 
and ${\hat \Gamma}^D$ is the matrix that 
diagonalizes the down-squark mass matrix (in the super-CKM basis).
Contrary to the chargino exchange, in this case 
there is no zeroth order term in the mass expansion 
and $C_A'$ is not negligible a priori: 
$[C'_A]^{Z^0}_{g} = [C_A]^{Z^0}_{g} (q_L\leftrightarrow q_R)$.
However, the bounds on down-type mass insertions obtained from 
$\Delta F=2$ processes are much stronger than in the up-type sector
(because of the presence of $\Delta F=2$ gluino-box amplitudes) \cite{GGMS}. 
Once these constraints (and the vacuum-stability ones) 
are taken into account, it is easy to realize 
that in all types of $d^i \to d^j$ transitions both
$[C'_A]^{Z^0}_{g}$ and $[C_A]^{Z^0}_{g}$ can barely
reach the magnitude of $|C_A^{\rm SM}|$.
This conclusion is important since it ensures that the 
model-independent constraints on $C_A$ discussed before 
cannot be invalidated by an accidental cancellation 
between $C_A'$ and $C_A$. 

\medskip
The whole discussion on axial-vector operators can thus be 
summarized as follows: (i) at present, in all types of $d^i \to d^j$ 
transitions, axial-vector contributions can 
reach the SM level at most; (ii) the most efficient way to 
constrain (or detect) non-standard axial-vector 
contributions is by means of non-helicity-suppressed 
decay modes.

\section{Scalar and pseudoscalar operators}
\label{sec:scalar}

As discussed in Refs.~\cite{Babu}--\cite{Nierste},
in the limit of large $\tan\beta$ Higgs-mediated 
scalar FCNCs can strongly modify, or even overcome, 
the SM amplitudes of rare helicity-suppressed modes.
The origin of this effect is related to the appearance, 
at the one-loop level, of an effective coupling between $H_U$ 
and down-type quarks \cite{HRS}. As discussed in \cite{IR} 
(see also \cite{Gian}),
a necessary condition to take into account all $\tan\beta$-enhanced 
terms in FCNC amplitudes is the diagonalization 
of the down-type quark mass matrix in the presence of these effective 
couplings. Here we generalize this approach to the
case where the effective $\bar d_R^id_L^jH_U^*$ coupling contains 
also non-minimal sources of flavour mixing.\footnote{~We 
stress that the so-called resummation of the 
$\tan\beta$-enhanced terms, in the limit where corrections
 of $O(M_W^2/{\tilde M}^2)$
are neglected, is nothing but the diagonalization
of the effective Yukawa interaction, which includes the dimension-four
effective operators appearing only 
at the quantum level \cite{IR,Gian}. In particular, 
the computation of the coefficients of all relevant 
dimension-four effective operators  to order $(\alpha_i/\pi)^1$, 
together with the diagonalization of the effective Yukawa interaction,
leads to a resummation of all terms of order $(\alpha_i \tan\beta/\pi)^n$
in the limit $M_W^2 \ll {\tilde M}^2$. }

Restricting the attention to the down-type neutral-Higgs sector, 
the effective down-type Yukawa interaction we need to consider 
is given by 
\beq
\cL_d^{\rm eff}  =  \dbar^i_R \left[  y^d_i \delta_{ij} H^0_D 
+ E_{ij} H_U^{0*} \right] d^j_L 
{\rm ~+~h.c.}~,
\label{eq:l_eff}
\eeq 
where $E_{ij}$ parametrizes the coupling generated at one loop
by both chargino and gluino diagrams. Here we do not make any 
specific assumption about the flavour structure of $E_{ij}$, 
but require that the off-diagonal entries can be treated as 
perturbations with respect to the (full) diagonal mass terms in 
(\ref{eq:l_eff}).\footnote{~In particular, this means we ignore 
the fine-tuned scenario where $E_{ii} \tan\beta \approx - y^d_i$.}
Then proceeding as in Ref.~\cite{IR}
with a perturbative diagonalization of the mass terms, we 
find 
\bea
\cL_{\rm FCNC}^{{\rm eff}~(i > j)} = \frac{y^d_i \tan\beta}{\bar y^d_i} 
\left[ E_{ij} + E_{ji}^* \left(\frac{\bar y^d_j}{\bar y^d_i} -\frac{y^d_j}{y^d_i} \right) 
\right] \left[ \frac{1}{\tan\beta} H_u^{0 *} - H_d^0 \right] {\bar d}_R^i d_L^j + 
{\rm h.c.}, \ \
\label{eq:LeffFC}
\eea
where 
\beq
\bar y^d_i =  y^d_i + E_{ii} \tan\beta~. 
\eeq
The analogue of (\ref{eq:LeffFC}) for $i<j$ is obtained with the exchange $i\leftrightarrow j$
in all indices of Yukawa couplings (both $y^d$ and $\bar y^d$), but not in the $E_{ij}$ terms.

In the absence of non-minimal sources of flavour mixing, one has
\beq
E_{ij} = E^g_{ij} + E^\chi_{ij}~, \qquad 
E^g_{ij} = \epsilon_g y^d_i \delta_{ij}, \qquad 
E^\chi_{ij} =  \epsilon_Y  y^d_i y_t^2 V^{0*}_{3i} V^0_{3j},  
\eeq
where  $V^0$ denotes the bare CKM matrix (without the $\epsilon_Y \tan\beta$ 
corrections induced by re-diagonalization) and
\bea
\epsilon_g = \frac{2\alpha_s}{3\pi} \frac{\mu M_g}{M^2_{{\tilde d}_L }}
f(x_{g{d}_L} , x_{{d}_R{d}_L} )\, , \qquad 
\epsilon_Y = \frac{1}{16\pi^2} \frac{\mu A}{M^2_{{\widetilde u}_L}}
f(x_{\mu {u}_L}, x_{{u}_R {u}_L})\, .
\label{eq:eY}
\eea
Using Eqs.~(\ref{eq:LeffFC})--(\ref{eq:eY}), it is easy to recover 
the results of Refs.~\cite{Babu,IR}. We shall now proceed 
with the evaluation of $E^g_{ij}$ and $E^\chi_{ij}$ in the presence 
of non-minimal sources of flavour mixing.

\subsection{Charginos}
\label{sec:Chi-H-penguin}
The $E_{ij}$ term induced by  chargino--up-squark loops can be written as
\be
E^\chi_{ij}= \frac{1}{16 \pi^2}L^j_{nl}{\bar R}^i_{nk} G_{nlk}~,
\ee
where 
\bea
{\bar R}^i_{nk}&=& y^d_i ({\hat \Gamma}^U)\da_{i_L k} {\hat U}\da_{2n},\\
G_{nlk}&=& y_t {\hat \Gamma}^U_{k q_L} (A_U V_0)\da_{q_L t_R}
({\hat \Gamma}^U)\da_{t_R l} \frac{M_{\chi_n}}{M^2_{{\tilde u}_k}} 
f(x_{lk},x_{nk})  \, .
\eea
As usual, the trilinear soft-breaking term has been decomposed 
as $A_U \times Y_U $, and we have neglected all up-type 
Yukawa couplings but the top one. Since $E_{ij} \propto y_i^d$,
only the case $i>j$ is relevant, and it is also clear that 
the $E_{ji}^*$ term in Eq.~(\ref{eq:LeffFC}) can safely 
be neglected. 

Integrating out the heavy neutral Higgs fields, in the 
large-$\tan\beta$ limit, we then obtain
\bea
[C_{S}]^{h^0,H^0,A^0}_{\chi}= [C_{P}]^{h^0,H^0,A^0}_{\chi}=
\frac{\pi^2}{2 G_F^2 M_W^2 M_A^2} \frac{ y_\ell E^\chi_{ij} \tan\beta}{
[ 1+ \tan\beta ( \epsilon_g  +  \epsilon_Y y_t^2 \delta_{i3})]}\, ,
\label{eq:C_S,P}
\eea
while the coefficients $[C'_{S,P}]^{h^0,H^0,A^0}_{\chi}$,
suppressed by the factor $y^d_j/y^d_i$,
turns out to be negligible. Due to the off-diagonal entries of the CKM matrix, 
even with a diagonal up-squark mass matrix $E^\chi_{i\not=j}\not=0$. This 
zeroth order term leads to \cite{IR}:
\beqa
&&[C_S]^{h^0 H^0 A^0}_\chi = [C_P]^{h^0 H^0 A^0}_\chi = \nnu\\
&&\quad\quad= \frac{ m_{d_i}  m_\ell m_t^2}{4 M_W^2}
\frac{ {\bar \lambda}^t_{ij} \tan^3\beta  }{
 [1+ \tan \beta ( \epsilon_g  +  \epsilon_Y y_t^2 \delta_{i3} )]  } 
\frac{\mu A f\left( x_{\mu {u}_L} , x_{{u}_R {u}_L} 
\right) }{ M_{\widetilde u_L}^2  M_A^2  [1+\epsilon_g \tan\beta] }~,
\label{eq:susy2}
\eeqa
where, expressing $V^0$ in terms of the physical CKM matrix,
\beq
{\bar \lambda}^t_{i > j} = \left\{ \ba{ll} V^*_{33}  V_{3j} & (i=3)~,  \\
 V^*_{3i}  V_{3j} \left[ \frac{ 1 + \tan \beta (\epsilon_g + \epsilon_Y y_t^2) 
 }{ 1 + \epsilon_g \tan \beta } \right]^2 \quad  &  (i\not =3)~. \ea \right.
\label{eq:barlam_def}
\eeq

Interestingly, this zeroth order term in the mass-insertion 
approximation (MIA), turns out to be the most interesting contribution 
induced by chargino exchange. Indeed, the only 
possibility to remove the CKM suppression factor $V_{3j}$ 
(the only one relevant in $B$ decays) is by means of a double mass 
insertion: one in the chargino line, the other in the $l$-squark
line. This leads to a potential enhancement factor 
$(\delta^U_{RL})_{3j} \delta^{^\chi}/V_{3j}$, which is partially
compensated by the suppression of the loop function with
two mass insertions, i.e.~a structure very similar 
to Eq.~(\ref{eq:deltachi}). Then, taking into account the bounds on 
$(\delta^U_{RL})_{3j}$ discussed in Section~\ref{sect:charg_1},
it turns out that this contribution can be at most 
of the same order as the zeroth-order term, 
or the minimal-flavour-violating (MFV) term. 

The suppression factor $V^*_{3i}$, which is relevant only to the 
kaon case, can be more easily removed: we simply need 
an off-diagonal trilinear term not proportional to $V^*_{3i}$. This leads to 
a net enhancement factor $(\delta^U_{LR})_{i3}/V_{3i}^*$,
which could reach one order of magnitude. However, as discussed in Ref.~\cite{IR},
the zeroth-order chargino contribution to the scalar 
$K_L \to \ell^+\ell^-$ amplitude amounts only to a few per cent of the SM one. 
Thus this potential enhancement is not particularly relevant. 

\subsection{Gluinos}
\label{sec:gluino-H-penguin}
The $E_{ij}$ term induced by gluino--down-squark loops can be written as
\be
E^g_{ij}= \frac{1}{16\pi^2} A^j_{bl}{\bar B}^i_{ak} J_{balk}~,
\ee
where 
\bea
{\bar B}^i_{ak} &=& {\bar A}^i_{ak} (i_L \rightarrow i_R)~, \\
J_{balk}&=& \mu \delta_{ab} {\hat \Gamma}^D_{k q_R} y^d_q   
({\hat \Gamma}^D)\da_{q_L l}
\frac{M_{g}}{M^2_{{\tilde d}_l}}
f(x_{g{d}_l}, x_{{d}_k{d}_l})~.
\eea
As can be noted, the gluino contribution to $E_{i\not=j}$
has a structure rather different from the chargino one.
The latter is different from zero already at the 
lowest order in the MIA, but it is necessarily proportional to 
the Yukawa coupling $y_i^d$. On the other hand, 
$E^g_{i\not=j}$ vanishes at the zeroth order in the MIA;
however, with a suitable mass insertion it 
can be made proportional to any of the down-type 
Yukawa couplings. 
As a result, in the gluino case also the  $C'_{S,P}$ 
coefficients can play an important role.

Assuming for simplicity $i>j$, the leading 
contributions to the scalar operators 
induced by  $E^g_{i\not=j}$ are given by
\bea
&& [C_S]^{h^0 H^0 A^0}_{g} = [C_P]^{h^0 H^0 A^0}_{g} = 
\frac{ 4 g_s^2}{ 3g^2}~
\frac{ m_{d_i} m_\ell\, \mu M_{\tilde g}}{ M_{A^0}^2 M^2_{\tilde d} }~
\frac{  \tan^3\beta}{
[1+\tan \beta (\epsilon_g+\epsilon_Y y_t^2 \delta_{i3}
)]^2} \times \nnu \\
&&\times
\left[  (\delta^D_{LL})_{ij}~
x_{d d_L}^2 f(x_{g d_L}, x_{ d_R d_L}, 1)
+ \frac{ y_b}{y_i^d}  ({\delta}^D_{RR})_{i3} ({\delta}^D_{LL})_{3j} ~
x^3_{dd_L} f( x_{g d_L}, x_{d_R d_L}, x_{d_R d_L}, 1)
\right],  \nnu \\
&& \label{eq:C_gluino} \\
&& [C'_{S}]^{h^0 H^0 A^0}_{g} = - [C'_{P}]^{h^0 H^0 A^0}_{g} =
\frac{ 4 g_s^2}{ 3g^2}~
\frac{ m_{d_i} m_\ell\, \mu M_{\tilde g}}{ M_{A^0}^2 M^2_{\tilde d} }~
\frac{  \tan^3\beta}{
[1+\tan \beta (\epsilon_g+\epsilon_Y y_t^2 \delta_{i3}
)]^2} \times \nnu \\
&& \times
\left[  (\delta^D_{RR})_{ij}~
x_{d d_R}^2 f(x_{g d_R}, x_{ d_L d_R}, 1)
+ \frac{ y_b}{y_i^d}  ({\delta}^D_{LL})_{i3} ({\delta}^D_{RR})_{3j} ~
x^3_{dd_R} f( x_{g d_R}, x_{d_L d_R}, x_{d_L d_R}, 1)
\right],  \nnu \\
&& \label{eq:C'_gluino}
\eea
where the off-diagonal elements in the squark mass matrix 
have been parametrized in terms of flavour-changing insertions in 
the sfermion propagators, defined as in Eq.~(\ref{eq:deltas}).
The terms arising at the second order in the MIA are relevant,
and play a very important role, only in the $s\to d$ case 
[$({\delta}^D_{LL(RR)})_{ii}\dot=0$]. Indeed, similarly to 
what happens in chargino-induced $Z$ penguins  \cite{Colangelo},
only with a double flavour-mixing of the type $(2\to 3)\times(3\to 1)$ 
can the large Yukawa coupling of the third generation contribute 
to $s\to d$ amplitudes. 

Before analysing the phenomenological implications of these formulae,
we point out some differences with respect to previous 
analyses:
\begin{itemize}
\item{} Contrary to Refs.~\cite{Chankowski,Bobeth2}, the expressions 
 (\ref{eq:C_gluino}) and  (\ref{eq:C'_gluino})
 contain the leading (flavour-diagonal) $\epsilon\tan\beta$ corrections
 to all orders.
 As pointed out in \cite{IR}, this is a numerically 
 important contribution for $\tan\beta \sim 50$, where the largest 
 effects are expected. 
 In the limit where higher-order $\epsilon\tan\beta$ corrections are 
 neglected, we find that $C_{S,P}^{g}$ is proportional to $\tan^3\beta$, similarly to 
 $C_{S,P}^{\chi}$. This behaviour, which is explicit in Ref.~\cite{Chankowski},
is somehow hidden in Eqs.~(4.4)--(4.5) of Ref.~\cite{Bobeth2}
(where the effective Yukawa interaction is not diagonalized). However, 
the $\tan^3\beta$  behaviour become manifest also in Ref.~\cite{Bobeth2} 
after an expansion of the ${\hat \Gamma}^D$ matrices:
in this limit we fully agree with the overall 
normalization of Ref.~\cite{Bobeth2}.\footnote{~This 
normalization differs from the one of  Ref.~\cite{Chankowski}, 
which contains an extra factor of 4. We thank J. Urban for a clarifying 
discussion about this point.} 
\item Within the limiting factors associated to the MIA, 
our formulae contain as a special case the scenario C of Ref.~\cite{Bobeth2}. 
In our opinion, such a scenario
should not be denoted as a case of {\em minimal} flavour violation.
Indeed diagonal --- but flavour-non-universal --- squark 
mass matrices do provide a {\em new} source of breaking of the 
$SU(3)^5$ flavour group~\cite{Georgi,Gian} and indeed they break 
the SM relation between Yukawa matrices and FCNC amplitudes. 
Since the purpose of this analysis is to identify the potentially 
largest effects generated by non-minimal sources of flavor mixing, 
and to identify the main differences with respect to MFV scenario
(defined as in Ref~\cite{Gian}), we prefer to treat all non-minimal
cases in the same way, using the MIA. For the same reason, we shall not consider 
subleading terms such as the neutralino amplitude, which become 
important only in presence of cancellations among the leading terms \cite{Bobeth2}.
\end{itemize}

\section{Phenomenological bounds}
As discussed in the previous two sections, only scalar (and pseudoscalar) 
operators can induce large enhancements of $B_{d,s}\rightarrow\ell^+\ell^-$ rates
over the SM expectations. Among the contributions 
to these operators, the one induced by chargino exchange is likely 
to be dominated by the MFV term, widely discussed in the literature. 
On the other hand, the gluino contribution is extremely sensitive to possible new sources 
of flavour mixing. For this reason, in the following we shall concentrate 
only on gluino-mediated scalar amplitudes.

\subsection{$B_{d,s} \rightarrow \mu^+ \mu^-$}
In the limit where $\tan\beta$ is large ($\tan\beta \gsim 30$),
$M_A$ is rather light ($M_A \lsim 500$~GeV) and the off-diagonal 
mixing terms $(\delta^D_{LL(RR)})_{3i}$ are not vanishing,  
the gluino (scalar) amplitude could dominate both 
$B_d\rightarrow\ell^+\ell^-$ and 
$B_s\rightarrow\ell^+\ell^-$ rates.
To get a rough idea of the possible effects, in the limit of 
degenerate sparticles we can write:
\bea
\frac{{\mathcal B}^{g}{(B_{d}\rightarrow \mu^+ \mu^-)}}{{\mathcal B}^{\rm SM}{(B_{d}\rightarrow \mu^+ \mu^-)}} 
&\approx & 3 \times 10^6 \left(\frac{200}{M_A} \right)^4  
\left|(\delta^D_{LL (RR)} )_{31}\right|^2 
\frac{\left(\frac{\tan\beta}{50} \right)^6}{\left[ 
\frac{2}{3} + \frac{1}{3} \left(\frac{\tan\beta}{50}\right) \right]^4}\, ,
\label{eq:DB_Bd}\\
\frac{{\mathcal B}^{g}{(B_{s}\rightarrow \mu^+ \mu^-)}}{{\mathcal B}^{\rm SM}{(B_{s}\rightarrow \mu^+ \mu^-)}} 
&\approx & 1.5 \times 10^5 \left(\frac{200}{M_A} \right)^4  
\left|(\delta^D_{LL (RR)} )_{32}\right|^2 
\frac{\left(\frac{\tan\beta}{50} \right)^6}{\left[ 
\frac{2}{3} + \frac{1}{3} \left(\frac{\tan\beta}{50}\right) \right]^4}\, .
\label{eq:DB_Bs}
\eea
These approximate numerical expressions hold at  
first order in the MIA, in the limit where we neglect any interference 
between supersymmetric and SM amplitudes and between supersymmetric 
contributions with different mass insertions. Moreover, we have neglected 
$\epsilon_Y y_t^2$ with respect to $\epsilon_g$ 
in the denominators of Eqs.~(\ref{eq:C_gluino}) and (\ref{eq:C'_gluino}),
and we have set $\epsilon_g$ to the value it assumes in the limit of degenerate 
supersymmetric scales, for $\mu>0$ ($\epsilon_g \approx 0.012$).\footnote{~The 
sign of $\mu$ is chosen according to the indications from 
$b\rightarrow s \gamma$ and $(g-2)_\mu$ \cite{g_2_mw}.}
A more complete illustration of the full dependence 
of ${\mathcal B}(B_{d}\rightarrow \mu^+ \mu^-)$ from various 
supersymmetric parameters is provided by Fig.~\ref{fig:1}.

\FIGURE[t]{
   \epsfig{file=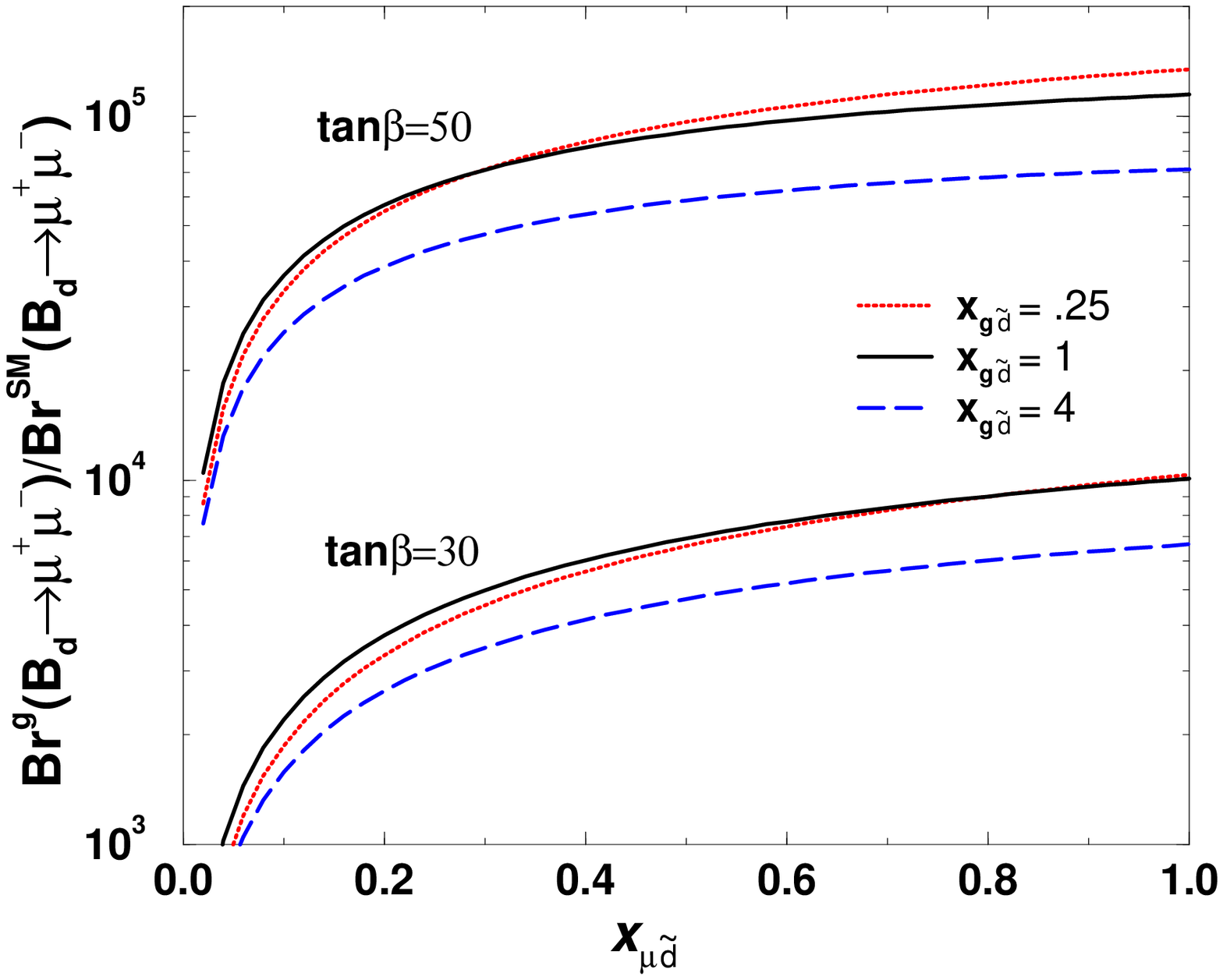,height=2.3in}  \hspace{0.1in} 
    \epsfig{file=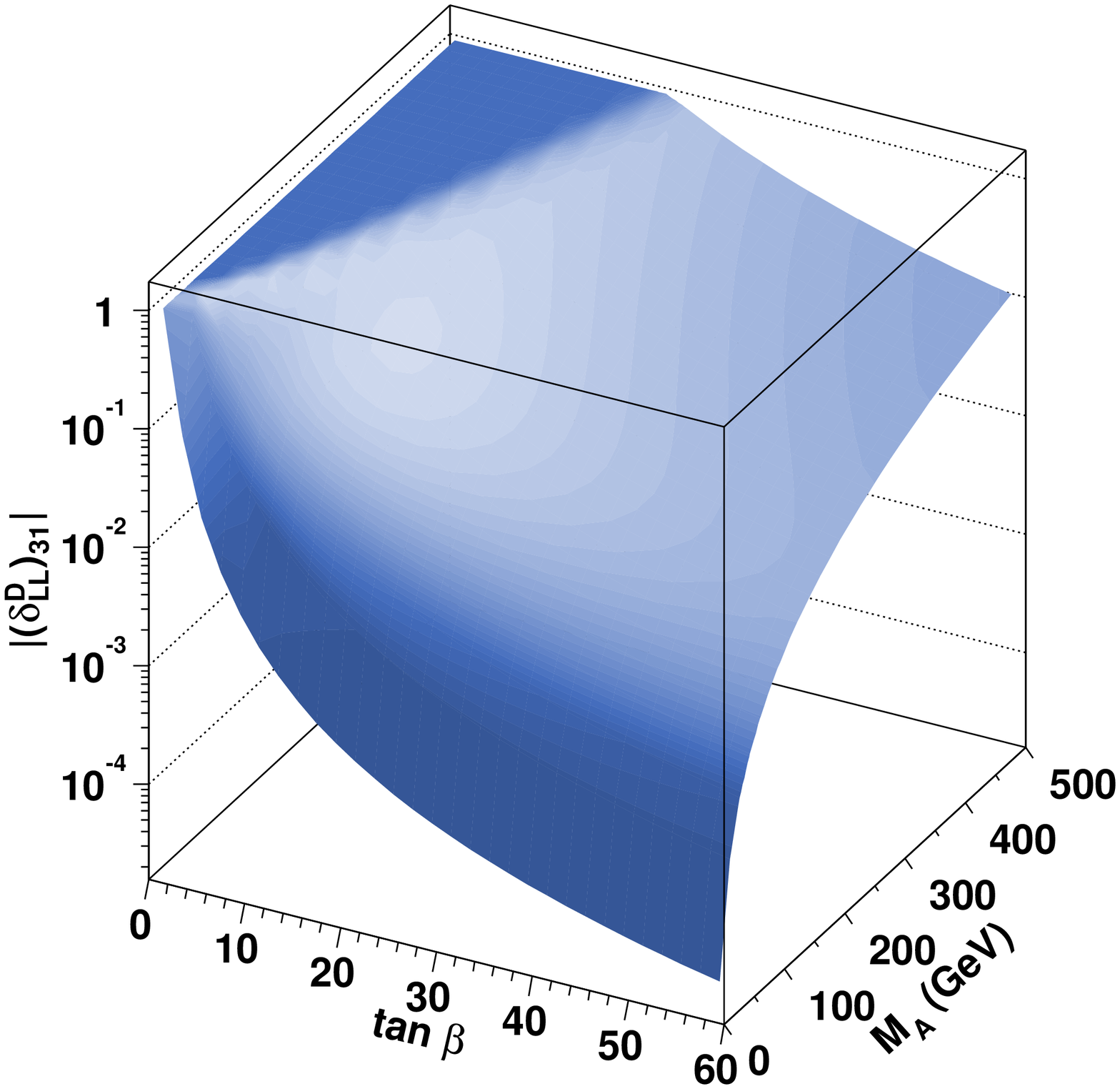,height=3.2in,width=2.45in} 
    \caption{Left: $\BR(B_{d}\rightarrow\mu^+\mu^-)$ normalized to the 
       SM prediction as a function of $x_{\mu{d}_L}$,      
       for $(\delta^D_{LL})_{31}=0.1$, $M_A=200~$GeV 
       and different values of $\tan\beta$ and $x_{g {d}_L}$.
       Right: present bounds on $|(\delta^D_{LL(RR)})_{31}|$
       imposed by the (scalar) gluino contribution to
       $\BR(B_{d}\rightarrow\mu^+\mu^-)$ 
       as a function of $\tan\beta$ and $M_A$, 
      for $x_{\mu {d}_L}=x_{{g}{d}_L}=1$.
\protect\label{fig:1} }}

Using the present experimental bounds
\bea
{\mathcal B}\,(B_d\rightarrow\mu^+\mu^-)&<&2.0\times10^{-7}\qquad  90\%~{\rm C.L.}~\cite{BaBar} \, ,
 \label{eq:BdExp} \\
{\mathcal B}\,(B_s\rightarrow\mu^+\mu^-)&<&2.0\times10^{-6}\qquad  90\%~{\rm C.L.} ~\cite{CDF}\, ,
 \label{eq:BsExp} 
\eea
we can derive a series of constraints on the squark flavour-mixing terms  
appearing in Eqs.~(\ref{eq:C_gluino}) and (\ref{eq:C'_gluino}).
These are reported in Table~\ref{tab:delta} and illustrated in 
Fig.~\ref{fig:1} (we show explicitly only the $1 \leftrightarrow 3$ case since the 
$2 \leftrightarrow 3$ is completely analogous, except for the 
rescaling of the SM contribution).
All the bounds have been obtained in the limit of  
complete degeneracy among the eigenvalues of down-type squark masses.
Moreover, we stress once more that these bounds are valid only in
the limit where the gluino scalar amplitude corresponding to a 
given flavour-mixing term dominates completely the decay 
rate. If $(\delta^D_{LL})_{3i}$ and  $(\delta^D_{RR})_{3i}$
are of the same order of magnitude, interference effects can no
longer be neglected and leads to $O(1)$ modifications of these 
bounds. For instance for  $(\delta^D_{LL})_{3i}=(\delta^D_{RR})_{3i}$
the bounds become more stringent by about a factor $\sqrt{2}$.

\TABLE[t]{
 \begin{tabular}{|c|c|c|c|c|c|}  \hline 
 \multicolumn{2}{|c|}{ }
 & \multicolumn{2}{c|}{$\vert(\delta^{D}_{LL(RR)})_{31}\vert$ }
& \multicolumn{2}{c|}{$\vert(\delta^{D}_{LL(RR)})_{32}\vert$}\\
\hline
$x_{\mu {d}_L}$ & $x_{g{d}_L}$ &$\tan\beta=30$  & 
$\tan\beta=50$ 
& $\tan\beta=30$  & $\tan\beta=50$  \\ 
   1&1  &0.089&0.026  &0.22&0.065\\
0.25&1  &0.13&0.035  &0.33&0.086\\
0.25&4  &0.16&0.043  &0.39&0.11\\
\hline
 \multicolumn{2}{|c|}{ }
 & \multicolumn{2}{c|}{$\vert \re (\delta^{D}_{LL(RR)})_{21}\vert$ }
& \multicolumn{2}{c|}{$\vert \re [(\delta^{D}_{RR(LL)})_{23}(\delta^{D}_{LL(RR)})_{31}]\vert$}\\
\hline
$x_{\mu {d}_L}$ & $x_{g{d}_L}$ &$\tan\beta=30$  & 
$\tan\beta=50$ 
& $\tan\beta=30$  & $\tan\beta=50$  \\ 
   1&1  &0.017&5.0$\times 10^{-3}$ &0.7$\times 10^{-3}$&2.0$\times 10^{-4}$\\
0.25&1  &0.025&6.6$\times 10^{-3}$ &1.0$\times 10^{-3}$&2.7$\times 10^{-4}$\\
0.25&4  &0.030&8.0$\times 10^{-3}$ &1.4$\times 10^{-3}$&3.7$\times 10^{-4}$\\
\hline 
 \end{tabular}
\caption{Present constraints on the off-diagonal 
squark mass terms that rule the (scalar) gluino 
contribution to $B_{s,d}\to \mu^+\mu^-$ (upper) and $K_L \to \mu^+\mu^-$ (lower).
All the bounds have been obtained for 
$M_A=200$ GeV and scale as $(M_A/200~{\rm GeV})^2$. 
\label{tab:delta} }
}

As already pointed out in Ref.~\cite{Chankowski}, for sufficiently large 
values of $\tan\beta$ the bounds on $|(\delta^D_{LL(RR)})_{31}|$
are more severe than those derived from the 
gluino-box contribution to  $B_d$--${\bar B}_d$ 
mixing \cite{team}. However,
we find that the effect is substantially reduced, 
with respect to what discussed in Ref.~\cite{Chankowski},
by the higher-order $\epsilon\tan\beta$ terms.

Already for $\tan\beta=30$, the upper bounds obtained on 
$|(\delta^D_{RR(LL)})_{32}|$ are much more severe 
than those that could be derived, in the near future, 
from  $B_s$--${\bar B}_s$ mixing
or the CP-violating asymmetry in $B_d\to \phi K_s$. 
As a result, if the scenario with a large 
$\tilde b_{R}$--$\tilde s_{R}$ mixing discussed in 
Ref.~\cite{CMM} is accompanied by a large value of $\tan\beta$,
which is a natural possibility in SO(10) models, 
the most interesting observables to look at are 
${\mathcal B}\,(B_{s,d}\rightarrow\mu^+\mu^-)$. 
The clear signature of this scenario would be 
$\BR(B_{s}\rightarrow\mu^+\mu^-)$ well above the 
SM expectation and, at the same time, a ratio 
$\BR(B_{d}\rightarrow\mu^+\mu^-)/\BR(B_{s}\rightarrow\mu^+\mu^-)$ 
different from $|V_{td}/V_{ts}|^2$ (i.e. the value expected 
in the absence of new flavour structures \cite{Gian}). 

\subsection{$K_L \rightarrow \mu^+ \mu^-$}

Within the SM the short-distance contribution to the 
$K_L\rightarrow\mu^+\mu^-$ branching ratio is predicted to be 
${\mathcal B}^{\rm SM}\,(K_L\rightarrow\mu^+\mu^-)_{\rm SD}= 
(8.7\pm3.7)\times 10^{-10}$ \cite{BBL}.
This expectation cannot be directly confronted with 
experimental data because of the long-distance amplitude 
generated by the two-photon intermediate state. 
Indeed the precise experimental measurement
of ${\mathcal B}^{SM}\,(K_L\rightarrow\mu^+\mu^-)$ 
\cite{Ambrose} turns out to be completely saturated by the 
absorptive two-photon contribution, signaling 
a cancellation between short-distance and 
dispersive long-distance amplitudes. 
Taking into account the estimates of the 
dispersive long-distance amplitude in Refs.~\cite{KL}, 
we shall assume in the following the conservative upper bound
\beq
{\mathcal B}\,(K_L\rightarrow\mu^+\mu^-)_{\rm SD}< 3.0\times 10^{-9}\, ,
\label{eq:KL_bound}
\eeq
which should be regarded as a reference figure rather than as
a strict bound. 

As can be read from Eqs.~(\ref{eq:C_gluino}) and (\ref{eq:C'_gluino}),
gluino contributions to the $K_L\rightarrow \ell^+ \ell^-$ amplitude
suffer, at the first order in the MIA, by a double strong suppression:
the smallness of the strange Yukawa coupling and the indirect constraints 
on $(\delta^D_{LL(RR)})_{21}$ \cite{team}. 
However, the SM amplitude of $K_L\to \ell^+ \ell^-$ decay is also 
affected by a strong suppression due to the CKM hierarchy
[$\cA(K_L\to \ell^+ \ell^-)^{\rm SM} \sim\lambda^5$]. 
As a result, for large values of $\tan\beta$ 
this supersymmetric amplitude could even become the 
dominant short-distance contribution.
Adopting the same approximations as in 
Eqs.~(\ref{eq:DB_Bd}) and (\ref{eq:DB_Bs}), we can write 
\bea
\frac{ {\mathcal B}^{g(1)}{(K_L\rightarrow \mu^+ 
\mu^-)}}{{\mathcal B}^{\rm SM}{(K_L
\rightarrow \mu^+ \mu^-)_{\rm SD}}} 
\approx \left(\frac{200}{M_A} \right)^4  \left| \frac{
\re (\delta^D_{LL (RR)} )_{21}  }{3\times 10^{-3}} \right|^2  
\frac{\left(\frac{\tan\beta}{50} \right)^6}{\left[ 
\frac{2}{3} + \frac{1}{3} \left(\frac{\tan\beta}{50}\right) \right]^4}.
\quad
\eea
If $|\re (\delta^D_{LL})_{12}|\gg | \im (\delta^D_{LL})_{12}|$, 
as happens in many reasonable scenarios, 
the bound on $|\re (\delta^D_{LL (RR)} )_{21}|$ 
we can extract from this amplitude (see Table~\ref{tab:delta}) 
becomes competitive, for $\tan\beta \sim 50$, with those derived 
from the gluino-box contribution to 
$K^0$--${\bar K}^0$ mixing~\cite{team}.

\FIGURE[t]{
   \epsfig{file=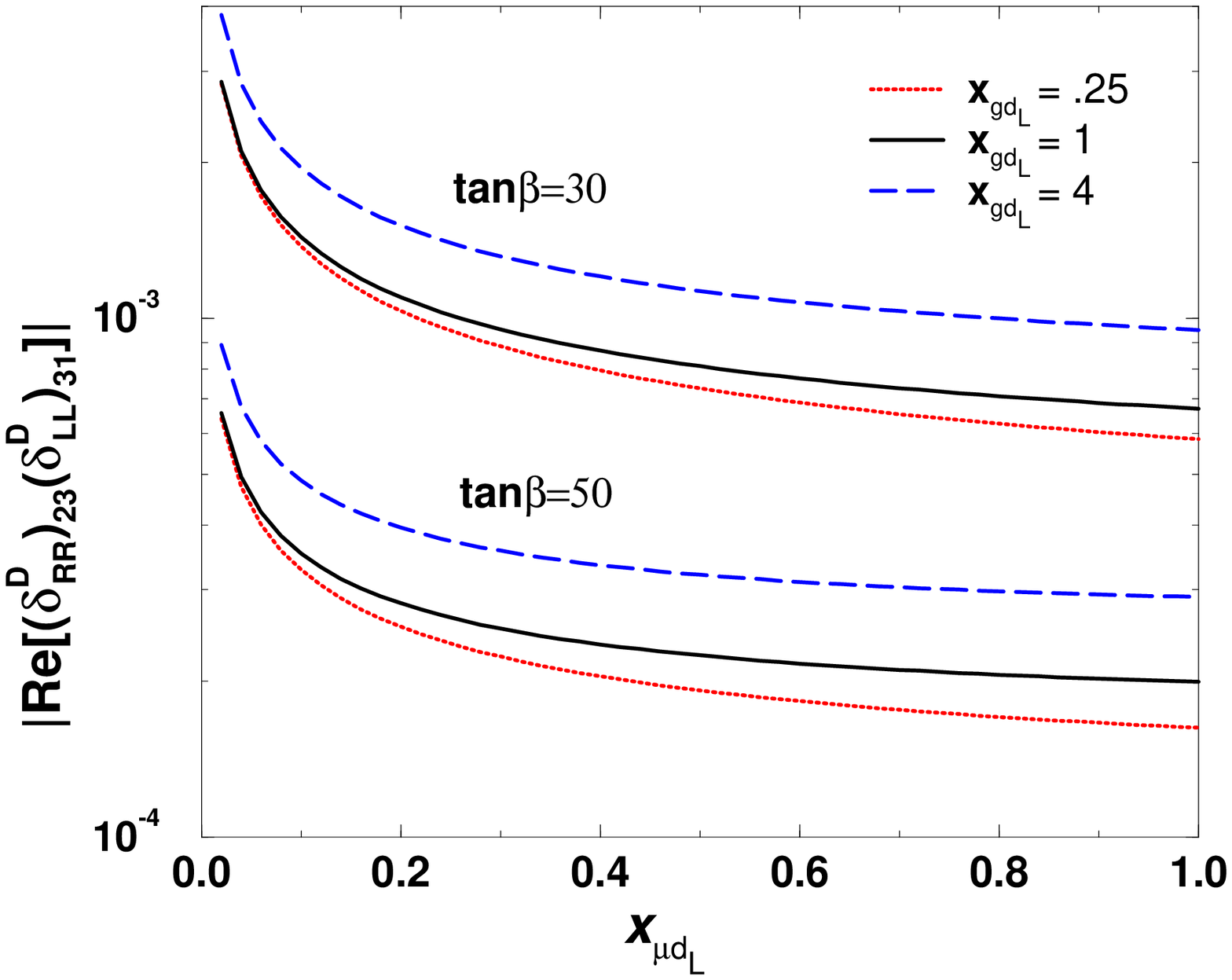,height=2.3in}  \hspace{0.1in} 
    \epsfig{file=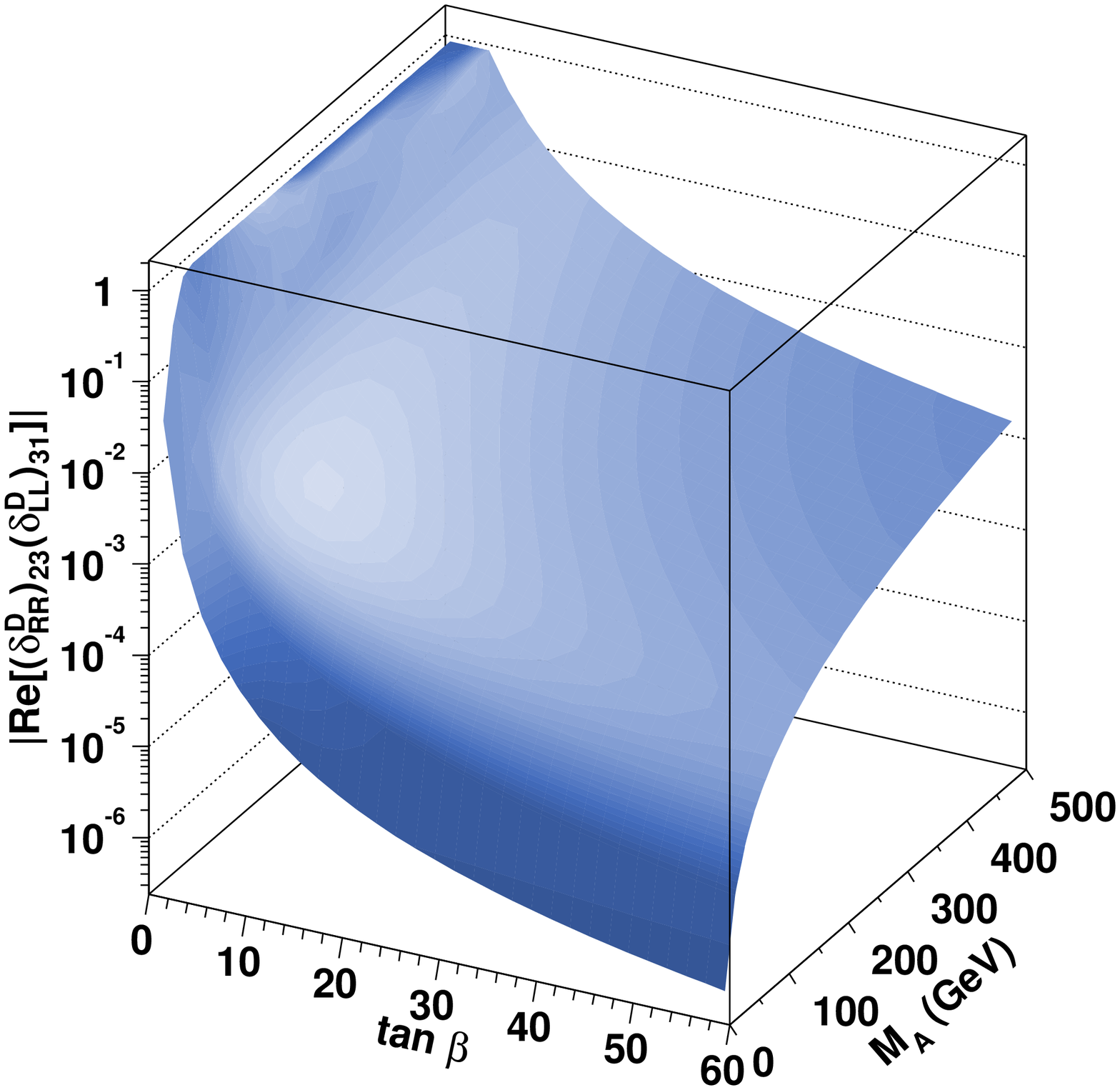,height=3.2in,width=2.45in} 
    \caption{Constraints on 
      $|\re [(\delta^D_{RR (LL)})_{23}(\delta^D_{LL (RR)})_{31}]|$ 
      set by $K_L \to \mu^+ \mu^-$. Left: the bound
       as a function of $x_{\mu {d}_L}$, for $M_A=200~$GeV 
       and different values of $\tan\beta$ and $x_{g {d}_L}$.
       Right: the bound  as a function of both $\tan\beta$ and $M_A$
      for $x_{\mu {d}_L}=x_{{g}{d}_L}=1$.   
\protect\label{fig:delta12}}}

As already emphasized, in rare $K$ decays it is often 
necessary to go up to the second order in the MIA 
in order to take into account all possible large effects 
\cite{Colangelo}. This occurs also in this case where,
according to Eqs.~(\ref{eq:C_gluino}) and (\ref{eq:C'_gluino}),
at the second order in the MIA we can get rid of both 
suppression factors affecting the term with a single 
mass insertion. In the usual limit of degenerate supersymmetric 
scales, we can write
\beqa
\frac{{\mathcal B}^{g(2)}{(
 K_L\rightarrow \mu^+ \mu^-)}}{{\mathcal B}^{\rm SM}{(K_L\rightarrow \mu^+ \mu^-)_{\rm SD}}} 
&\approx& \left(\frac{200}{M_A} \right)^4  
\left| \frac{ \re [(\delta^D_{RR (LL)} )_{23}(\delta^D_{LL (RR)})_{31}] 
 }{ 1 \times 10^{-4} } \right|^2 
\frac{\left(\frac{\tan\beta}{50} \right)^6}{\left[ 
\frac{2}{3} + \frac{1}{3} \left(\frac{\tan\beta}{50}\right) \right]^4}\,. \nnu \\
\eeqa
Notice that the value of the effective coupling 
$|\re [(\delta^D_{RR(LL)})_{23}(\delta^D_{LL(RR)})_{31}]|$
needed to reach the SM level is one order of magnitude 
smaller than the corresponding one required by 
$| \re (\delta^D_{LL})_{12}|$.
The precise bounds imposed by Eq.~(\ref{eq:KL_bound})
on the effective coupling 
$|\re [(\delta^D_{RR(LL)})_{23}(\delta^D_{LL(RR)})_{31}]|$,
for different values of supersymmetric parameters,
are shown in  Fig.~\ref{fig:delta12} and reported 
in Table~\ref{tab:delta}. These bounds indicate that
if $(\delta^D_{LL})_{ij}$ and $(\delta^D_{RR})_{ij}$ are of the 
same order, we cannot saturate at the same time the present limits 
on $\BR(B_{s}\rightarrow \mu^+ \mu^-)$ and $\BR(B_{d}\rightarrow \mu^+ \mu^-)$.
On the other hand, this possibility remains open if 
there exists a strong hierarchy between the two types of 
couplings. As we shall see, this conclusion is strongly 
reinforced by the constraints from $B$--$\bar B$ mixing.

\subsection{$B$--$\bar B$ mixing}
So far we compared the bounds on the mass insertions relevant
to $\Delta F=1$ scalar amplitudes, at large $\tan\beta$, with
constraints imposed by $\Delta F=2$ vector-type amplitudes
(the gluino-box diagrams), which are independent from  $\tan\beta$.
However, as pointed out first in \cite{Buras_2P} (see also \cite{Buras_last}), 
at large $\tan\beta$ scalar amplitudes could also play 
a significant role in $\Delta F=2$ transitions. Indeed 
contracting the complex $H_d$ field in Eq.~(\ref{eq:LeffFC}) 
with an appropriate FCNC coupling of $H_d^*$ [hidden in the 
Hermitian conjugate part of Eq.~(\ref{eq:LeffFC})] we obtain  
$\Delta F=2$ effective operators of the type 
${\bar d}_R^i d_L^j {\bar d}_L^i d_R^j$
whose coupling scales like $\tan^4\beta$. 

If $E_{ij}$ is necessarily proportional to $y^d_i$,
as happens in the chargino case, it is easy to realize 
that these $\Delta F=2$ scalar operators are always suppressed 
by a small Yukawa coupling ($y^d_{j\not=3}$) \cite{Buras_2P}. 
A similar situation occurs in the gluino case,
if we allow only one type of mass insertion
to be different from zero, i.e. $(\delta^D_{LL})_{3j}$ 
or $(\delta^D_{RR})_{3j}$. 
In these cases the only non-trivial constraints are set by 
$B_s$--$\bar B_s$ mixing, where this small Yukawa coupling is $y_s$.
We have explicitly checked that these bounds 
are less stringent than those set by $\BR(B_{s}\rightarrow\mu^+\mu^-)$,
i.e. they do not exclude the possibility that 
$\BR(B_{s}\rightarrow\mu^+\mu^-)$ is just below
its present experimental limit.\footnote{~A similar conclusion was 
recently reported also in Ref.~\cite{Rosiek2}.}

\TABLE[t]{
 \begin{tabular}{|c|c|c|c|c|c|}  \hline
 \multicolumn{2}{|c|}{ } 
 & \multicolumn{2}{c|}{$\sqrt{\vert (\delta^{D}_{RR})_{31}(\delta^{D}_{LL})_{31} \vert} $}
& \multicolumn{2}{c|}{$ \sqrt{\vert (\delta^{D}_{RR})_{32}(\delta^{D}_{LL})_{32} \vert} $}\\
\hline
$x_{\mu {d}_L}$ & $x_{g{d}_L}$ &$\tan\beta=30$  & 
$\tan\beta=50$ 
& $\tan\beta=30$  & $\tan\beta=50$  \\ 
   1&1  &0.005  &2.3$\times 10^{-3}$ & 0.020 & 0.009 \\
0.25&1  &0.010  &4.7$\times 10^{-3}$ & 0.039 & 0.019 \\
0.25&4  &0.011  &5.4$\times 10^{-3}$ & 0.045 & 0.022 \\
\hline 
 \end{tabular}
\caption{Present constraints on the off-diagonal 
squark mass terms which contribute, via scalar amplitudes,
to $\Delta M_{B_d}$ and $\Delta M_{B_s}$.
All the bounds have been obtained for 
$M_A=$ $200$ GeV and scale linearly with $M_A$. 
\label{tab:delta2} }
}

On the other hand, as we learned from the 
$P \rightarrow \ell^+ \ell^-$ analysis, 
if we allow both LL and RR insertions to be different from zero we 
can have both $E^g_{3j}$ and $E^g_{j3}$ proportional to $y_b$ and 
get rid of any small Yukawa coupling. In this case we obtain the 
following $\Delta F=2$ effective Hamiltonian: 
\beqa
\cH^{\rm eff~(scalar)}_{\Delta B=2} &=& - \frac{ E^g_{3j} E^{g*}_{j3} \tan^2\beta }{ 
  M_A^2 [ 1 + \tan \beta (\epsilon_g + \epsilon_Y y_t^2) ]^2 }\,   {\bar b}_R d_L^j {\bar b}_L d_R^j \nnu \\
 &=&  -\frac{y_b^2 \tan^2\beta (\delta^D_{LL})_{3j} (\delta^D_{RR})_{3j}  }{ 
         [ 1 + \tan \beta (\epsilon_g + \epsilon_Y y_t^2) ]^2 M_A^2} 
 \left(\frac{2\alpha_s \mu M_{\tilde g} }{3\pi M_{\tilde d}^2 }\right)^2  \, {\bar b}_R d_L^j {\bar b}_L d_R^j  \nnu \\
 &&  \qquad\qquad 
     \times \left[ x_{d d_L}^2 x_{d d_R}^2 f(x_{g d_L}, x_{ d_R d_L}, 1) f(x_{g d_R}, x_{ d_L d_R}, 1) \right]~,
     \qquad 
\eeqa
which have a large potential impact on both $\Delta M_{B_d}$
and $\Delta M_{B_s}$. Employing the usual approximation of degenerate
supersymmetric scales, we can write
\bea
 \left| \frac{ \Delta M^{g}_{B_{d}} }{   \Delta M^{\rm SM}_{B_{d}} } \right| 
&\approx &  \left(\frac{200}{M_A} \right)^2  
\left| \frac{ (\delta^D_{LL})_{31}(\delta^D_{RR})_{31} }{ 6 \times 10^{-6} } \right|
\frac{\left(\frac{\tan\beta}{50} \right)^4}{\left[ 
\frac{2}{3} + \frac{1}{3} \left(\frac{\tan\beta}{50}\right) \right]^4}\, ,
\label{eq:DMB_Bd}\\
 \left| \frac{ \Delta M^{g}_{B_{s}} }{   \Delta M^{\rm SM}_{B_{s}} } \right| 
&\approx & \left(\frac{200}{M_A} \right)^2  
\left| \frac{ (\delta^D_{LL})_{32}(\delta^D_{RR})_{32} }{ 1 \times 10^{-4} } \right|
\frac{\left(\frac{\tan\beta}{50} \right)^4}{\left[ 
\frac{2}{3} + \frac{1}{3} \left(\frac{\tan\beta}{50}\right) \right]^4}\, ,
\label{eq:DMB_Bs}
\eea
The bounds obtained by the full analytical expressions, 
imposing the conditions $| \Delta M^{g}_{B_{d,s}} | <
|  \Delta M^{\rm SM}_{B_{d,s}} |$, are reported in 
Table~\ref{tab:delta2}. The comparison between 
Table~\ref{tab:delta} and Table~\ref{tab:delta2} clearly
shows that down-type squark mixing could 
have a sizeable impact on $\BR(B_{s,d}\rightarrow \mu^+ \mu^-)$
modes only in the presence of a strong hierarchy 
between  LL and RR insertions.

We finally mention that, in principle, one could 
extract bounds on the combination 
$(\delta^D_{LL})_{23}(\delta^D_{LL})_{31}(\delta^D_{RR})_{23}(\delta^D_{RR})_{31} $ 
from $K$--$\bar K$ mixing. We have explicitly 
checked that the bounds imposed by $\Delta M_K$
on $\sqrt{| \re [(\delta^D_{LL})_{23}(\delta^D_{LL})_{31}(\delta^D_{RR})_{23}(\delta^D_{RR})_{31}   ]|}$
are slightly less stringent than those obtained from $K_L \to \mu^+\mu^-$ 
on $|\re [(\delta^D_{LL})_{23}(\delta^D_{RR})_{31}]|$ and 
reported in Table~\ref{tab:delta}.
This happens because, contrary to $B_{s,d}\to \mu^+\mu^-$ modes, 
we can allow at most $O(1)$ effects in $K_L \to \mu^+\mu^-$.

\section{Conclusions}
The rare dilepton decays $B_{s,d} \to \ell^+\ell^-$
offer a unique opportunity to study
the scalar FCNCs which naturally arise in supersymmetry at 
large $\tan\beta$. In this paper we have complemented previous 
studies of these decays modes in supersymmetry 
\cite{Babu}--\cite{Nierste} analysing their sensitivity to 
non-minimal sources of flavour mixing. 

Similarly to the MFV scenario \cite{Babu,IR},
we have shown that also in this case it is necessary to 
diagonalize the effective Yukawa interaction ---
including the $\bar d_R^id_L^jH_U^*$ coupling
generated by the new flavour structures --- in order to 
identify the correct $\tan\beta$ behaviour of
scalar FCNCs and to take into account 
all the leading $\tan\beta$-enhanced terms. 
Using this approach, we have corrected/completed previous 
results on $\cA(B_{s,d} \to \ell^+\ell^-)$ in the presence 
of non-minimal sources of flavour mixing \cite{Chankowski,Bobeth2}.

An important difference with respect to the MFV scenario arises 
because the effective  $\bar d_R^id_L^jH_U^*$ operator generated
by gluino--squark loops is not necessarily proportional to the 
Yukawa coupling of the right-handed external quark 
($y^d_i$): allowing both left--left and right--right mass 
insertions to be different from zero, we can generate 
potentially large contributions suppressed only by 
the bottom Yukawa coupling (even for both $i\not=3$ and $j\not=3$).
This leads to extract stringent bounds on the flavour 
off-diagonal down-type squark mixing, at large $\tan\beta$,
from $\Delta M_{B_{d,s}}$ and $K_L \to \mu^+\mu^-$.
These bounds, which have been analysed for the first 
time as reported in this paper, are summarized in Tables~1 and 2.

As a result of the bounds from  $\Delta M_{B_{d,s}}$ and $K_L \to \mu^+\mu^-$,
we reached the conclusion that flavour off-diagonal 
squark mixing can induce order-of-magnitude enhancements 
of $\BR(B_{s,d} \to \ell^+\ell^-)$ only in the presence of a 
strong hierarchy between left--left and right--right 
flavour-mixing terms. This scenario, which is apparently fine-tuned,
could be generated in a natural way in GUT models, as pointed out for 
instance in Ref.~\cite{CMM}.  If this is the case, 
one should expect a ratio 
$\BR(B_{d}\rightarrow\mu^+\mu^-)/\BR(B_{s}\rightarrow\mu^+\mu^-)$ 
very different from $|V_{td}/V_{ts}|^2$, i.e. the value expected 
in the absence of new flavour structures. 

\section*{Acknowledgements}
We thank A. Dedes and J. Urban for useful discussions.

\appendix
\section{Loop functions}
The loop functions appearing in the chargino--squark and gluino--squark 
penguin amplitudes discussed in Sections~\ref{sec:axial}  
and \ref{sec:scalar}, 
can be derived from the recursive formula~\cite{Kpnn1,Colangelo}
\beq
I(x,y,z_1,\ldots,z_{n-2})=
\frac{I(x,z_1,\ldots,z_{n-2})-I(y,z_1,\ldots,z_{n-2})}{x-y}~,
\eeq
starting from the single variable functions
\bea
f(x)= \frac{x\log x}{x-1} \qquad\qquad  {\rm and }\qquad  \qquad k(x)= x f(x)\, .
\eea
Some useful reference values are 
\bea
f(1,1)=\frac{1}{2}~,\qquad 
f(1,1,1)=-\frac{1}{6}~,\qquad 
f(1,1,1,1)=\frac{1}{12}~.
\eea

\end{document}